\documentclass{elsart}

\usepackage{amssymb}
\usepackage{hyperref}

\begin{document}
\begin{frontmatter}


%

\title{Periodic systems in time: double-well potential}

\author{Sergei~P.~Maydanyuk}
\ead{maidan@kinr.kiev.ua, Sergei.Maydanyuk@fuw.edu.pl}
\address{
Institute for Nuclear Research,
National Academy of Sciences of Ukraine, \\
47, prosp.~Nauki, Kiev-28, 03680, Ukraine
}

\begin{abstract}
Time analysis of oscillations of a particle between wells
in the one-dimensional double-well potential with infinite
high outside walls, based on wave packet use and energy 
spectrum analysis, is presented.
For the double-well potential of the form $x^{2}+1/x^{2}$ in 
the external regions, an exact analytical solution of the 
energy spectrum is found (by standard QM approach), an analysis 
of oscillation periodicity is fulfilled, an approach for exact
analitical calculation of the oscillation period is proposed
(for the first time).
\end{abstract}

\begin{keyword}
double-well potential   \sep
discrete spectrum       \sep
oscillation period      \sep
wave packet             \sep
exactly solvable model  \sep
periodic system         \sep
tunneling time          \sep
supersymmetry

\PACS
03.65.-w \sep      
03.65.Db \sep      
03.65.Ge \sep      
03.65.Xp \sep      
\end{keyword}
\end{frontmatter}

\section{Introduction}

A quantum system, representing a particle in the potential field
in the form of two wells with a finite high barrier and infinite
high outside walls, is used in many different tasks of
physics and chemistry \cite{Selg.2000.PS}.

If the tunneling of the particle through the barrier in such
system is possible then there are transitions of this particle
between the wells, named as \emph{``oscillations''}.
In the general case, the oscillations are not periodic (and also
harmonic) motion. For their description one can introduce the
following characteristics:
\begin{itemize}

\item
the time duration, after which the system returns into its
initial state (this time characteristic can be named as
\emph{oscillation period of the particle between the wells}
\cite{Maydanyuk.1999.JPS}) or \emph{period of Poincare's cycle});

\item
the time duration, after which the system has passed into a
state closed as much as possible with initial one.
\end{itemize}

Instanton methods \cite{Coleman.1988.NUPHA,Zhou.2000.PHLTA} are
powerful tools for realization of a time analysis of a behaviour
of the particle in the double-well potential.
An alternative and less widespread approach for the time
analysis of the double-well system behaviour consists in
use of wave packets in spite of the fact that such system has
discrete energy spectrum only \cite{Razavy.1988.PRPLC}.
If to localize the packet in the initial time moment
inside one well than one can suppose that a maximum of this
packet has passed into another well after some time duration.
One can define the oscillation period of the particle between
wells for the systems with a periodic motion and also one can
define the time duration of the most probable return into the
initial state for the system, which motion is not periodic,
on the basis of transition of the packet maximum or the center
of mass of this packet between the wells.
Note that the packet maximum and the center of mass of the
packet describe the tunneling process of the packet through the
barrier in different ways.
Here, if in the tunneling the localization of the wave packet
in the space region of the barrier is improbable
(apparently, it is possible only in one case of essential
contribution into the total packet of its component for
sub-barrier energies),
than a motion of the center of mass of the packet in the
barrier region looks natural enough.

An approach for calculation of these time characteristics for
the double-well systems on the basis of analysis of energy
spectrum is presented in this paper.

\section{An analysis of the time periodicity of wave function
on the basis of the energy spectrum}

Let's consider a system with the discrete energy spectrum
where energy levels are located one from another at such 
distances, for which one can calculate exactly the largest
general divisor:
\begin{equation}
  E_{n} = E_{0} + \Delta \cdot n,
\label{eq.2.1}
\end{equation}                                  
where $n \in 0, N$ ($N$ is a set of natural numbers).
In particular, a harmonic oscillator, a particle inside box,
a charged particle in the Coulomb field satisfy to this
condition. Wave function of such system has a form \cite{Shiff}:
\begin{equation}
\Psi(x, t) =
  e^{-iE_{0}t/\hbar}
  \sum_{n} g_{n} \psi_{n}(x) e^{-i(E_{n}-E_{0})t/\hbar} =
  e^{-iE_{0}t/\hbar}
  \sum_{n} g_{n} \psi_{n}(x) e^{-i\Delta n t/\hbar},
\label{eq.2.2}
\end{equation}                                  
where $\psi_{n}(x)$ are the orthonormal functions of the
stationary states of the system satisfied to equation 
$\stackrel{\land}{H} \psi_{n}(x) = E_{n} \psi_{n}(x)$;
$\stackrel{\land}{H}$ is Hamiltonian of the system;
$\sum\limits_{n} |g_{n}|^{2} = 1$.
We choose the time moment $t = 0$ for time zero.

Consider the function
\begin{equation}
  f(x, t) = \Psi(x,t) e^{iE_{0}t/\hbar}
\label{eq.2.3}
\end{equation}                                  
in the time interval
$[-\frac{\pi\hbar}{\Delta}, \frac{\pi\hbar}{\Delta}]$.
It satisfy to Dirichle's condition \cite{Vorobiev}:
a) one can divide this interval into a finite number of intervals
where the function $f(x,t)$ is continuous, monotonous and
finite;
b) if $t_{0}$ is a point of discontinuity of the function
$f(x,t)$, then $f(x, t_{0}+0)$ and $f(x, t_{0}-0)$ exist. And
the sum $\sum_{n} g_{n} \psi_{n}(x) e^{-i(E_{n}-E_{0})t/\hbar}$
in (\ref{eq.2.2}) is an expansion of the function $f(x,t)$ into
the Fourier trigonometric series (with constant coefficients
$\Delta/\hbar$) by variable $t$. Such series is convergent at
any point $t$ of the interval
$[-\frac{\pi\hbar}{\Delta}, \frac{\pi\hbar}{\Delta}]$,
where the function $f(x,t)$ is continuous in $t$. In any break
point $t_{0}$ the Fourier series converges to
$\frac{f(x, t_{0}+0) + f(x, t_{0}-0)}{2}$.
Therefore, the function $f(x,t)$ is periodic by variable $t$
in the interval considered above. This periodic dependence
can be extended into whole range of definition of this function
by $t$. One can calculate the period by such a way:
\begin{equation}
  T_{f} = \frac{2\pi\hbar}{\Delta}.
\label{eq.2.4}
\end{equation}                                  
An exact analytical dependence of the periodic function
$f(x,t)$ on time variable $t$ is determined by the total set
of eigenfunctions $\psi_{n}(x)$ with coefficients $g_{n}$ at
any point $x$ and is changed in dependence on $x$.

We find out an influence of the ground state with the level $E_{0}$
on a periodicity of the function $\Psi(x,t)$ in time. Consider
the following example. Let the function $f(x_{0},t)$ at point
$x=x_{0}$ be harmonic by variable $t$ (assume that this function
can be written as $const \cdot e^{-i\Delta t/\hbar}$), then we
obtain:
\begin{equation}
  \Psi(x_{0},t) =
  \mbox{const} \cdot e^{-iE_{0}t/\hbar} e^{-i\Delta t/\hbar} =
  \mbox{const} \cdot e^{-i(E_{0}+\Delta)t/\hbar}.
\label{eq.2.5}
\end{equation}                                  
One can see that the function $\Psi(x_{0},t)$ is harmonic and
periodic by time variable $t$ also. We find an exact analytical
solution for a period of the function $\Psi(x_{0},t)$ with
taking into account the first level $E_{0}$:
\begin{equation}
  T_{\Psi} = \frac{2\pi\hbar}{E_{0}+\Delta} =
  \frac{2\pi\hbar}{E_{1}}.
\label{eq.2.6}
\end{equation}                                  
Using (\ref{eq.2.6}), we can analyse a contribution of the first
level $E_{0}$ in the period $T_{\Psi}$.

In particular, if to use the parabolic function
$U(x) = \frac{mw^{2}x^{2}}{2}$ as the potential energy of the
system (where $m$ is a mass of the particle, $w$ is an
oscillation frequency of the particle in classical mechanics),
then the energy spectrum of the system has a form:
\begin{equation}
\begin{array}{ll}
  E_{n} = (n + \frac{1}{2}) \hbar w, &
  n = 0, 1, 2 ...
\end{array}
\label{eq.2.7}
\end{equation}                                  
If there is a coordinate $x_{0}$ where the function $f(x_{0},t)$
is harmonic by time variable $t$ then the oscillation period
relatively this coordinate decreases in $1,5$ times with taking
into account the first level $E_{0} = \frac{\hbar w}{2}$
(i.~e. the contribution of the first level is not small):
\begin{equation}
\begin{array}{ll}
  T_{old} =
  \displaystyle\frac{2\pi\hbar}{\Delta} =
  \displaystyle\frac{2\pi\hbar}{\hbar w}, &
  T_{new} =
  \displaystyle\frac{2\pi\hbar}{E_{0}+\Delta} =
  \displaystyle\frac{2\pi\hbar}{(\frac{1}{2}+1) \hbar w} =
  \displaystyle\frac{2}{3} \cdot \frac{2\pi\hbar}{\hbar w} =
  \displaystyle\frac{2}{3} T_{old}.
\end{array}
\label{eq.2.8}
\end{equation}                                  

Therefore, the function $\Psi(x,t)$ is harmonic in time
at point $x$ only in such case when at this point $x$
the function $f(x,t)$ is harmonic in time also.

If the function $f(x,t)$ is not harmonic then one can consider
the function $\Psi(x,t)$ as periodic in time approximately
and evaluate its period. Let $f_{1}(t)$ and $f_{2}(t)$ be two
periodic functions with periods $T_{1}$ and $T_{2}$, accordingly.
We assume that these functions are harmonic in enough small
neighborhood of the point ($x$, $t$).
The period $T$ (i.~e. the time duration, after which the system
has passed into a state closed as much as possible with initial
one) can be calculated from the following relation:
\begin{equation}
  \displaystyle\frac{1}{T} =
  \displaystyle\frac{1}{T_{1}} + \displaystyle\frac{1}{T_{2}}.
\label{eq.2.10}
\end{equation}                                  
Taking into account that the functions $f(x,t)$ and
$\exp{(-iE_{0}t/\hbar)}$ are periodic by variable $t$, one can
determine conditions of the periodicity of the function
$\Psi(x,t)$.

Any quantum system with discrete energy spectrum performs
a finite motion. But if there is the largest divisor $\Delta$
determined exactly for energy level of this system and the
condition (\ref{eq.2.1}) is satisfied, then in general case
one can consider the time behaviour of this system as periodic
approximately and one can calculate the period more
accurately by (\ref{eq.2.6}) or (\ref{eq.2.10})
(i.~e. with taking into account of the first level)
whereas the expression (\ref{eq.2.4})
(i.~e. without taking into account the first level)
is less accurate. The calculation accuracy of oscillation
period can be estimated on the basis of the set of orthogonal
functions $\psi_{n}(x)$ with coefficients $g_{n}$ at selected
point $x$.
This method of the estimation of the periodicity of the wave
function can be used for the time analysis of the oscillations
of the particle between wells in the double-well potential with
infinite high outside walls.

\section{Exactly solvable models with symmetric double-well
potential}

Apparently, the spectrum of the double-well potential with any
form is not equidistant. Nevertheless, we find the double-well
potential form, for which there is the exact analytical
dependence of the wave function on time and one can analyse
its periodicity on the basis of the energy spectrum.

Let's consider a quantum system, representing a particle in the
symmetric double-well potential field of a form:
\begin{equation}
  U(x) = \displaystyle\frac{mw^{2}}{2}
  \left\{
  \begin{array}{ll}
    (x + x_{0})^{2} +
    \displaystyle\frac{B^{2}}{(x + x_{0})^{2}}, &
    \mbox{for } x>a>0;          
    \\
     C - D x^{2},               &
     \mbox{for } -a<x<a;        
     \\
    (x - x_{0})^{2} +
    \displaystyle\frac{B^{2}}{(x - x_{0})^{2}},  &
    \mbox{for } x<-a<0;          
\end{array}
\right.
\label{eq.3.1}
\end{equation}                                  
where $m > 0$, $w > 0$, $B > 0$, $C > 0$ and $D > 0$. $C$
determines the barrier height. We assume that $B$, $C$ and $D$
have such values that U(x) is continuous at points $x = \pm a$.
Minimums of two wells are located at points
$x_{1}=-\sqrt{B}+x_{0}$ and $x_{2}=\sqrt{B}-x_{0}$.

The potential (\ref{eq.3.1}) is symmetric. In result of
sub-barrier tunneling and above-barrier propagation there are
transitions of the particle between wells, i.e. its oscillations.
One can analyse the period of such oscillations on the basis of
the energy spectrum. For finding the spectrum it is need to
resolve the stationary Schr\"{o}dinger equation
and take into account all conditions, to which the wave function
satisfies. In the calculation of the energy spectrum for the
potential (\ref{eq.3.1}) the boundary conditions at
$x \to -\infty$ and $x \to +\infty$ play the important role.
According with analysis, the conditions of the continuity of the
wave function and its derivative at points $x=\pm a$ do not differ
the localization of the energy levels.
Therefore, the solution of the stationary Schr\"{o}dinger
equation in the external regions $x<-a$ and $x>a$ has caused
the main interest in calculation of the energy spectrum.

\subsection{Calculations of the energy spectrum}

We use new parameters:
\begin{equation}
\begin{array}{lcl}
  G = \displaystyle\frac{2mE}{\hbar^{2}},               &
  F =-\displaystyle\frac{m^{2}w^{2}}{\hbar^{2}},        &
  K =-\displaystyle\frac{m^{2}w^{2}B^{2}}{\hbar^{2}}.
\end{array}
\label{eq.3.2}
\end{equation}                                          

Then the stationary Schr\"{o}dinger equation in the external
regions transforms into a form:
\begin{equation}
  \displaystyle\frac{d^{2} \psi}{d\bar{x}^{2}} +
  \left(G + F\bar{x}^{2} +
  \displaystyle\frac{K}{\bar{x}^{2}}\right) \psi = 0,
\label{eq.3.3}
\end{equation}                                          
where
\begin{equation}
\begin{array}{ll}
  \bar{x} = x-x_{0}, & \mbox{for } \bar{x}<-a; \\
  \bar{x} = x+x_{0}, & \mbox{for } \bar{x}>a.
\end{array}
\label{eq.3.4}
\end{equation}                                          

We find a solution of this equation. Fulfill the replacements:
\begin{equation}
\begin{array}{lcl}
  \xi       = \alpha \bar{x}^{2},     &
  \alpha    = \displaystyle\frac{mw}{\hbar} = \sqrt{-F},        &
  \psi(\xi) = \biggl(\displaystyle\frac{\xi}{\alpha}\biggr)^{-1/4}
              w(\xi).
\end{array}
\label{eq.3.5}
\end{equation}                                          

Then the equation (\ref{eq.3.3}) transforms into the standard
Whittaker form \cite{Abramowitz}:
\begin{equation}
  \displaystyle\frac{d^{2}w}{d\xi^{2}} + \biggl[ -\displaystyle\frac{1}{4} +
  \displaystyle\frac{G}{4\sqrt{-F}} \displaystyle\frac{1}{\xi} +
  \biggl( \displaystyle\frac{3}{16} + \displaystyle\frac{K}{4} \biggr)
  \displaystyle\frac{1}{\xi^{2}} \biggr] w = 0.
\label{eq.3.6}
\end{equation}                                  

Include the following parameters
\begin{equation}
\begin{array}{llll}
   k    = \displaystyle\frac{G}{4 \sqrt{-F}}, &
\mu^{2} = \displaystyle\frac{1}{16} - \displaystyle\frac{K}{4}, &
   a    = \displaystyle\frac{1}{2} - k + \mu, &
   c    = 1 + 2 \mu
\end{array}
\label{eq.3.7}
\end{equation}                                  
and fulfill the following replacement
\begin{equation}
 y(\xi) = \xi^{-c/2}e^{\xi/2}w(\xi).
\label{eq.3.8}
\end{equation}                                  
Then the equation (\ref{eq.3.6}) transforms into the
hypergeometric equation of the form:
\begin{equation}
  \xi \displaystyle\frac{d^{2}y}{d\xi^{2}} + (c - \xi)
  \displaystyle\frac{dy}{d\xi} - ay = 0.
\label{eq.3.9}
\end{equation}                                  
Its partial solutions can be presented using the hypergeometric
function $F(a,c;\xi)$ in a form:
\begin{equation}
\begin{array}{rcl}
y_{1}(\xi) & = & F(a,c;\xi), \\
y_{2}(\xi) & = & \xi^{1-c} F(a-c+1,2-c;\xi), \\
y_{3}(\xi) & = & e^{\xi} F(c-a,c;-\xi), \\
y_{4}(\xi) & = & \xi^{1-c} e^{\xi} F(1-a,2-c;-\xi).
\end{array}
\label{eq.3.10}
\end{equation}                                  

Consider the case: $c \not\in Z$. Here, the solutions $y_{3}$
and $y_{4}$ can be written through $y_{1}$ and $y_{2}$ by
Kummer's transformation \cite{Abramowitz}. Therefore, they
depend linearly on $y_{1}$ and $y_{2}$. We obtain the first
two solutions $y_{1}$ and $y_{2}$ with initial variables:
\begin{equation}
\begin{array}{l}
\psi_{1}(\bar{x}) = \alpha^{1/2 + \mu}
        \bar{x}^{-1+2\mu} e^{-\alpha \bar{x}^{2} / 2}
        F\left(a, 1+2\mu; \alpha \bar{x}^{2}\right), \\
\psi_{2}(\bar{x}) = \alpha^{1/2 + \mu}
        \bar{x}^{-1} e^{-\alpha \bar{x}^{2} / 2}
        F\left(a-2\mu, 1-2\mu; \alpha \bar{x}^{2} \right).
\end{array}
\label{eq.3.11}
\end{equation}                                  

In accordance with the finiteness condition of the wave function,
one can write:
\begin{equation}
\begin{array}{l}
\mbox{for } \psi_{1}(x) : a \in 0, - N; 2 \mu \not\in - N, \\
\mbox{for } \psi_{2}(x) : -a+2\mu \in 0,N; 2 \mu \not\in N.
\end{array}
\label{eq.3.12}
\end{equation}                                  
After satisfying of these conditions the series
(which define the hypergeometric functions for the solutions
$\psi_{1}$ and $\psi_{2}$) transform into polynomial, and the
energy spectrum becomes discrete.
One can find from the expressions (\ref{eq.3.12}) that the
solutions $\psi_{1}$ and $\psi_{2}$ can not be used at the same
time. But both $\psi_{1}$ and $\psi_{2}$ corresponds to the
same energy spectrum.
Write expressions for the energy spectrum and wave function
(we point out the wave function in the external regions only):
\begin{equation}
\left\{
\begin{array}{ll}
  E_{n}^{-} = 2\hbar w \left(\displaystyle\frac{1}{2} + n - \mu\right); \\
  \psi_{1,n}(x) = \alpha^{1/2 - \mu}
                \bar{x}^{-1-2\mu}e^{-\alpha \bar{x}^{2}/2}
                F(-n, 1-2\mu; \alpha \bar{x}^{2}), &
                \mbox{for $x<-a$ and $x>a$};   \\
  \psi_{2,n}(x) = \alpha^{1/2 + \mu}
                \bar{x}^{-1}e^{-\alpha \bar{x}^{2}/2}
                F(-n, 1-2\mu; \alpha \bar{x}^{2}), &
                \mbox{for $x<-a$ and $x>a$};
\end{array}
\right.
\label{eq.3.13}
\end{equation}                                          
\begin{equation}
\left\{
\begin{array}{ll}
  E_{n}^{+} = 2\hbar w \left(\displaystyle\frac{1}{2} + n + \mu\right); \\
  \psi_{1,n}(x) = \alpha^{1/2 + \mu}
                \bar{x}^{-1+2\mu}e^{-\alpha \bar{x}^{2}/2}
                F(-n, 1+2\mu; \alpha \bar{x}^{2}), &
                \mbox{for $x<-a$ and $x>a$}; \\
  \psi_{2,n}(x) = \alpha^{1/2 - \mu}
                \bar{x}^{-1}e^{-\alpha \bar{x}^{2}/2}
                F(-n, 1+2\mu; \alpha \bar{x}^{2}), &
                \mbox{for $x<-a$ and $x>a$}.
\end{array}
\right.
\label{eq.3.14}
\end{equation}                                          
Here
\begin{equation}
\begin{array}{ll}
\mu = \displaystyle\frac{1}{4}\sqrt{1 +
       \displaystyle\frac{4m^{2}w^{2}B^{2}}{\hbar^{2}}}, &
\bar{x} =
  \left\{
  \begin{array}{ll}
    |x - x_{0}|,  &
    \mbox{for } x<-a;   \\
    |x + x_{0}|,  &
    \mbox{for } x>a,
\end{array}
\right.
\end{array}
\label{eq.3.15}
\end{equation}                                  
$n \in 0, N$ ($N$ is the natural number set)
and the total set of the levels $E_{n}$ consists from
sets $E_{n}^{+}$ and $E_{n}^{-}$.

One can consider a co-existence of two eigenfunctions $\psi_{1}$
and $\psi_{2}$ (not dependent linearly one from another) for
any level $E_{n}$ as the doubly degeneracy of the energy
spectrum. In similar cases, an additional quantum number is used
for marking the difference between such states.
Inclusion of the asymmetry in such double-well potential gives
the double splitting of the energy spectrum and leads to
degeneracy removal.
Instanton methods give a similar result.

\subsection{The time analysis of the particle oscillations
between the wells}

We write the expressions (\ref{eq.3.13}) and (\ref{eq.3.14})
for the energy spectrum by such a way:
\begin{equation}
\begin{array}{l}
  E_{n}^{\pm} =
  2\hbar w \left(\displaystyle\frac{1}{2} \pm \mu\right)
  + 2\hbar w n =
  E_{0}^{\pm} + \Delta \cdot n, \\
  E_{0}^{\pm} =
  2\hbar w \left(\displaystyle\frac{1}{2} \pm \mu\right), \\
  \Delta = 2\hbar w.
\end{array}
\label{eq.3.16}
\end{equation}                                  
From here one can conclude that two sets of the energy levels
$E_{n}^{+}$ and $E_{n}^{-}$ represent independently equidistant
spectrum. One can study them separately, as described two
independent waves with their periods. Without of taking into
account the first levels $E_{0}^{+}$ and $E_{0}^{-}$, the
periods for these waves are equal:
\begin{equation}
  T_{f} =
  \displaystyle\frac{2\pi\hbar}{\Delta} =
  \displaystyle\frac{\pi}{w}.
\label{eq.3.17}
\end{equation}                                  

With taking into account the first levels $E_{0}^{+}$ and
$E_{0}^{-}$, a dependence of the wave function $\Psi(x,t)$ on
time is defined by such a way:
\begin{equation}
  \Psi(x,t) = e^{-iE_{0}^{+}t/\hbar} f^{+}(x,t) +
  e^{-iE_{0}^{-}t/\hbar} f^{-}(x,t),
\label{eq.3.18}
\end{equation}                                  
where $f^{+}(x,t)$ and $f^{-}(x,t)$ are the wave functions of
these waves without of consideration of the levels $E_{0}^{+}$
and $E_{0}^{-}$.
The expression (\ref{eq.3.18}) represents the exact analytical
dependence of the wave function on time.
One can consider approximately these two items as periodic with
the periods $T^{+}$ and $T^{-}$ calculated on the basis of
(\ref{eq.2.6}):
\begin{equation}
\begin{array}{ll}
  T^{+} = \displaystyle\frac{2\pi\hbar}{E_{0}^{+}+\Delta} =
  \frac{2\pi\hbar}{E_{1}^{+}}, &
  T^{-} = \displaystyle\frac{2\pi\hbar}{E_{0}^{-}+\Delta} =
  \frac{2\pi\hbar}{E_{1}^{-}}. \\
\end{array}
\label{eq.3.19}
\end{equation}                                  

One can calculate approximately the period of the particle
oscillation between the wells in the potential (\ref{eq.3.1}):
\begin{equation}
  \displaystyle\frac{1}{T} =
  \displaystyle\frac{1}{T^{+}+T^{-}} +
  \displaystyle\frac{1}{T^{+}-T^{-}}.
\label{eq.3.20}
\end{equation}                                  

In accordance with (\ref{eq.3.16}), a distance between two
nearest levels $E_{n}^{+}$ and $E_{n}^{-}$ (considered in some
tasks as a splitting value of energy spectrum $E_{n}$ in
result of tunneling through the barrier) can be calculated by
such a way:
\begin{equation}
\Delta E = E_{n}^{+} - E_{n}^{-} = \hbar w \sqrt{1 +
           \displaystyle\frac{4m^{2}w{2}}{\hbar^{2}} B^{2}}.
\label{eq.4.1}
\end{equation}                                  
If the barrier form satisfies to conditions of use of
semi-classical methods then one can find a dependence of the
penetrability coefficient $D$ of the barrier on the oscillation
period $T$ and on the largest divisor $\Delta$
\cite{Maydanyuk.1999.JPS}:
\begin{eqnarray}
D & \sim & \pi^{2} \biggl( 1 + \displaystyle\frac{4m^{2} w^{2}}
  {\hbar^{2}} B^{2} \biggr) =
  \pi^{2} \biggl( 1 + \displaystyle\frac{m^{2} B^{2}} {\hbar^{4}}
  \Delta^{2} \biggr) = \nonumber \\
  &  =   & \pi^{2} \biggl( 1 + \displaystyle\frac{\pi^{2}16 m^{2}
  B^{2}}{\hbar^{2}} \displaystyle\frac{1}{T^{2}} \biggr).
\label{eq.4.2}
\end{eqnarray}                                  

\section{Conclusions and perspectives}

The approach for the time analysis of the double-well systems on
the basis of energy spectrum is presented in this paper. For the
double-well symmetric potential with the form $x^{2} + B^{2}/x^{2}$
in the external regions the energy spectrum is calculated exactly
analytically and the analysis of the time periodicity is fulfilled.
The approach to approximate calculation of the oscillation period
is described. Similar potentials were studied in \cite{Otchik}.

During last two decades an essential progress in study of quantum
system properties has achieved after rapid development of methods
of supersymmetry in their application to quantum mechanics
(here, one can note a review \cite{Cooper.1995.PRPLC},
which should be the best SUSY QM review at opinion of the author).
Application of such methods (in particular, see last subsections
in \cite{Khare.math-ph/0409003}) to the analysis of periodicity of
the particle motion (oscillations) between two wells in the
double-well potential looks perspective and interesting in future
study of periodical systems.
For example, a question about change of the periodicity
characteristics of the particle oscillations after going to new
isospectral potentials from the potential considered in this paper
can be interesting.

For the double-well non-periodic quantum system one can select
``quasi-cycles'' (after which the system has passed into a state
closed as much as possible with initial one) with needed accuracy
\cite{Maydanyuk.1999.JPS}.
The maximal values of wave function of such systems are localized
inside a finite space region and one can calculate the oscillation
period with needed accuracy (with taking into account the needed
number of ``quasi-cycles'').
From the other side, it is interesting to use such approach in
the generalization of the SUSY QM methods, developed for the
analysis of the periodical quantum systems (i.~e. for the analysis
of \emph{quasi-periodical} systems).


\bibliographystyle{h-elsevier3}


\newpage
\begin{center}
{\large \bf Address:}
\end{center}

{\bf Sergei~P.~Maydanyuk} \\

Institute for Nuclear Research,
National Academy of Sciences of Ukraine,

47, prosp.~Nauki, Kiev-28, 03680, Ukraine \\

E-mail: {\em maidan@kinr.kiev.ua, Sergei.Maydanyuk@fuw.edu.pl}

Phone:  {\em (380-44) 265-46-92, 265-23-49}

Fax.:   {\em (380-44) 265-44-63}

\end{document}